\documentclass[twocolumn,preprintnumbers,amsmath,amssymb,showpacs]{revtex4}
\usepackage{epsfig}
\begin{document}
\title{How to get from imaginary to real chemical potential}
\author{Felix Karbstein\footnote{Electronic address: felix@theorie3.physik.uni-erlangen.de}}
\author{Michael Thies\footnote{Electronic address: thies@theorie3.physik.uni-erlangen.de}}
\affiliation{Institut f\"ur Theoretische Physik III,
Universit\"at Erlangen-N\"urnberg, D-91058 Erlangen, Germany}
\date{\today}
\begin{abstract}
Using the exactly solvable Gross-Neveu model as theoretical laboratory, we analyse in detail
the relationship between a relativistic quantum field theory at real and imaginary chemical
potential. We find that one can retrieve the full information about the phase diagram of the theory
from an imaginary chemical potential calculation. The prerequisite is to evaluate and analytically
continue the effective potential for the chiral order parameter, rather than thermodynamic
observables or phase boundaries. In the case of an inhomogeneous phase, one needs to compute
the full effective action, a functional of the space-dependent order parameter, at imaginary chemical
potential. 
\end{abstract}
\maketitle
\section{Introduction}
Progress in understanding the phase diagram of quantum chromodynamics (QCD) at finite temperature and
chemical potential is hampered by the sign problem: A real chemical potential yields in general a complex fermion
determinant, thereby rendering the standard Monte-Carlo algorithms inapplicable. One interesting idea to overcome
this problem is to compute at imaginary chemical potential where the sign problem is absent \cite{1,2,3,4}. 
An imaginary chemical potential is primarily a formal trick. Nevertheless, it has two physical interpretations
we are aware of:

1) The grand canonical partition function
\begin{equation}
{\cal Z}_{\rm gc}(T,\mu) = {\rm Tr}\, {\rm e}^{-\beta(\hat{H}-\mu \hat{N})} \qquad  (\beta=1/T)
\label{A1}
\end{equation}
at imaginary $\mu$ can be viewed as Fourier transform of the
canonical partition function
\begin{equation}
{\cal Z}_{\rm c}(T,N) = {\rm Tr}\left( {\rm e}^{-\beta \hat{H}}\delta(\hat{N}-N)\right)
\label{A2}
\end{equation}
with respect to fermion number,
\begin{equation}
{\cal Z}_{\rm c}(T,N) = \int_{-\pi}^{\pi}\frac{{\rm d}\theta}{2\pi} {\rm e}^{-{\rm i}N\theta}
{\cal Z}_{\rm gc}(T,\mu={\rm i}\theta/\beta ).
\label{A3}
\end{equation}
Since the canonical partition function yields any thermodynamic observable, it seems
at first sight that the sign problem has been overcome.
Unfortunately, this point of view is rather academic. 
A numerical Fourier transform for large volumes and particle numbers is out of question, as it would require
exponentially increasing accuracy. Although this method has been applied successfully
to small systems \cite{2,4a}, the thermodynamic limit cannot be reached. The only known way of
evaluating integrals with extremely rapidly varying integrand is the saddle point method. The
saddle point is the solution of 
\begin{equation}
\frac{\partial}{\partial \theta} \left\{ {\rm i} \theta N + \beta \Psi(T,{\rm i}\theta/\beta) \right\} =0
\label{A4}
\end{equation}
with $\Psi$ the grand canonical potential,
\begin{equation}
\Psi (T,\mu)=- \frac{1}{\beta}\ln {\cal Z}_{\rm gc}(T,\mu)
\label{A4a}
\end{equation}
Since $\Psi$ is an even function of $\mu$ or $\theta$ due to CP invariance, Eq.~(\ref{A4})
cannot have a real solution for $\theta$. Thus evaluation of the integrand at the saddle point takes us back to
the original problem of chemical potential with a real part, and nothing has been gained. 

2) A second physical interpretation arises if we invoke the so-called temperature inversion symmetry \cite{5,6,7,8}.
Due to covariance, the partition function at inverse temperature $\beta$ for a system with one
compact space dimension is invariant under exchange of space and imaginary time, provided
one uses periodic (anti-periodic) boundary conditions for bosons (fermions),
\begin{equation}
{\cal Z}(\beta,L) = {\cal Z}(L,\beta).
\label{A5}
\end{equation}
If one takes the limit $L\to \infty$, one can relate a hot, extended system to a cold, compressed system,
e.g. the free energy is related to the Casimir (ground state) energy. An imaginary chemical potential
can be ``gauged away" by a space-dependent phase
transformation of the fermion fields ($L=\beta$), 
\begin{equation}
\psi'(x) = {\rm e}^{{\rm i}\theta x/L} \psi(x).
\label{A6}
\end{equation} 
Like a magnetic field in an extra dimension, this changes the boundary conditions in the compact direction
from anti-periodic to quasi-periodic ones,
\begin{equation}
\psi'(L)=-{\rm e}^{{\rm i}\theta} \psi'(0)
\label{A7}
\end{equation}
interpolating between fermionic ($\theta=0$) and bosonic ($\theta=\pi$) 
boundary conditions. Phase transitions as a function of imaginary chemical potential in the original system 
go over into quantum phase transitions as a function of the boundary condition (or the fictitious magnetic field).

The 2nd point of view gives a physical picture, but no clue as to how to get from imaginary to real $\mu$.
The first one suggests such a pathway which however is illusionary. In practical lattice calculations,
the following strategy has usually been adopted \cite{3,4,9,9a,10,10a,10b}: 
One tries to determine observables or phase boundaries by analytic continuation
from imaginary to real $\mu$. Typically, one evaluates some function of $\mu$ at imaginary $\mu$, fits it
by a low order polynomial (or a ratio of polynomials \cite{10c,10d}) and then uses it at real chemical potential. The main drawback
here are the limitations due to singularities in the complex plane. Moreover, there does not seem to be 
any way to follow a phase boundary beyond a tricritical point, so that the phase diagram cannot be fully 
reconstructed, no matter how much effort is invested at imaginary $\mu$.

In the case of QCD, very little is known about the analytic structure of the thermodynamic potential
in the complex $\mu$ plane. Besides, even calculations at imaginary $\mu$ suffer from limited accuracy and
numerical noise.
In order to gain more insight into the problem of how to use imaginary $\mu$-computations most efficiently to learn
about the real world, we propose to look at a simple, solvable model field theory. We choose the Gross-Neveu (GN)
model in 1+1 dimensions \cite{11} where in the large $N$ limit all thermodynamic observables can easily be computed 
at both real and imaginary chemical potential, to any desired accuracy. Nevertheless the phase diagram
at real $\mu$ is non-trivial, sharing qualitative properties with QCD (see \cite{12} for a recent review).
We then pretend that we are only able 
to compute all quantities at imaginary $\mu$ and try to extrapolate them to real $\mu$, controlling our procedure
against the exact results at each step. In this way, we are able to come up with a scheme which allows to
overcome some of the limitations mentioned above and suggests some future work in lattice QCD.

This paper is organized as follows. In Sect. II, we briefly consider free massless fermions in 1+1 dimensions where 
everything can be done analytically. This serves to illustrate the general principles in a fortunate case
where there is no phase transition and all analytic continuations can be done in closed form. In Sect. III, we consider
massive free fermions. After these preparations, we turn to the GN model in Sect. IV. We apply the common
procedure of analytically continuing the phase boundary and show how to go beyond that. We restrict 
ourselves to the translationally invariant solution with homogeneous condensates.
In Sect. V we briefly contemplate the recently found inhomogeneous phase (soliton crystal) of the GN model 
and investigate whether one could find it at all with imaginary $\mu$. In Sect. VI we summarize our findings
and draw some conclusions concerning lattice QCD calculations.

\section{Free massless fermions in 1+1 dimensions}
Our starting point is the 
grand canonical (gc) potential density at real chemical potential, regularized via an UV momentum cutoff,
\begin{eqnarray}
\psi &=& - \frac{1}{\beta \pi} \int_0^{\Lambda/2} {\rm d}k \ln \left[ \left(1+{\rm e}^{-\beta(k-\mu)}\right)\left(
1+{\rm e}^{\beta(k+\mu)}\right)\right]
\nonumber \\
&=& - \frac{\Lambda^2}{8\pi} - \frac{\Lambda \mu}{2\pi} \label{B1} \\
& &  - \frac{1}{\beta \pi} \int_0^{\infty} {\rm d}k \ln \left[ \left(1+{\rm e}^{-\beta(k-\mu)}\right)\left(
1+{\rm e}^{-\beta(k+\mu)}\right)\right]
\nonumber
\end{eqnarray}
As usual we drop the divergent vacuum terms, i.e., use the bottom line of Eq.~(\ref{B1}) as definition
of the regularized gc potential density. The integral over momenta yields the familiar result
\begin{equation}
\psi = - \frac{\pi}{6\beta^2} - \frac{\mu^2}{2\pi}.
\label{B2}
\end{equation}
For later use we also compute the free energy density by a Legendre transformation,
\begin{eqnarray}
f & = & \psi-\mu \frac{\partial \psi}{\partial \mu} 
\nonumber \\
& = & - \frac{\pi}{6\beta^2} + \frac{\pi}{2} \rho^2
\label{B3}
\end{eqnarray}
with the fermion density
\begin{equation}
\rho=- \frac{\partial \psi}{\partial \mu} = \frac{\mu}{\pi}. 
\label{B4}
\end{equation}
We now turn to the less familiar gc potential density at imaginary chemical potential $\mu={\rm i} \theta/\beta$.
It can be expressed in closed form with the help of the dilogarithm, 
\begin{eqnarray}
\psi &=&  - \frac{1}{\beta \pi} \int_0^{\infty} {\rm d}k \ln \left[ \left(1+{\rm e}^{-\beta k+ {\rm i}\theta}\right)\left(
1+{\rm e}^{-\beta k- {\rm i}\theta}\right)\right]
\nonumber \\
& = & \frac{1}{\pi \beta^2} \left[ {\rm dilog} \left(1+{\rm e}^{{\rm i}\theta}\right) +  {\rm dilog} \left(1+{\rm e}^{-{\rm i}\theta}\right)
\right]
\label{B5}
\end{eqnarray}
[we use the Maple notation ${\rm dilog}(z)={\rm Li}_2(1-z)$]. $\psi$ is a periodic function in $\theta$
which can be simplified to
\begin{equation}
\psi = - \frac{\pi}{6\beta^2} + \frac{\theta^2}{2\pi \beta^2}
\label{B6}
\end{equation}
in the interval $[-\pi, \pi]$.
\begin{figure}
\epsfig{file=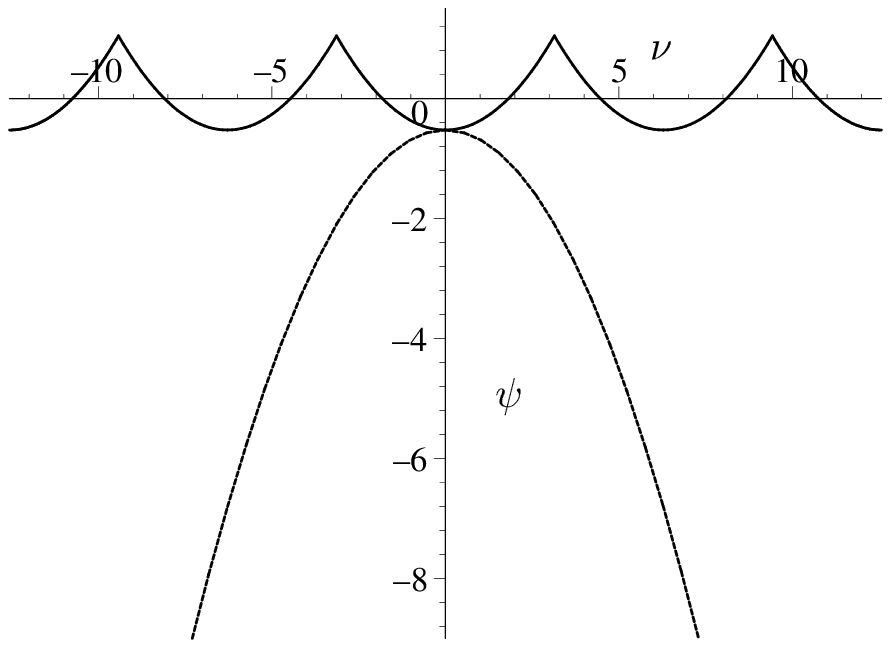,width=5cm,angle=270}
\caption{Grand canonical potential density for massless fermions. Lower parabola: real, upper periodic curve:
imaginary chemical potential ($\nu=\beta \mu={\rm i}\theta$). Units are such that $\beta=1$.}
\end{figure}
Fig.~1 shows the comparison between $\psi$ at real and imaginary chemical potential.
The discontinuities in the first derivative at $\theta=(2n+1)\pi$ ($n \in {\rm Z}$) is due to the branch point of ${\rm Li}_2(z)$ at
$z=1$.
A power series expansion in $\theta$ around  $\theta=0$ reduces to a second order
polynomial for $|\theta|<\pi$. It is noteworthy that the singularity at $\pi$ can be given a physical interpretation: At this value,
the Fermi-Dirac distribution is converted into a Bose-Einstein distribution at zero chemical potential. The singularity
is the same one which gives rise to Bose-Einstein condensation in higher dimensions.

The fact that everything can be done explicitly in this simple example enables us to illustrate the general ideas
mentioned in the Introduction.
We first verify the claim that the gc partition function at imaginary chemical potential
can be used in principle to derive the canonical partition function by projecting out a fixed fermion number,
see Eq.~(\ref{A3}). The size of the system will be denoted by $L$. For large enough $L$ we may extend the integration
limits to $\pm \infty$ so that we merely have to perform a Gaussian integral
\begin{eqnarray}
{\rm e}^{-\beta L f} & \approx &  \int_{-\infty}^{\infty} \frac{{\rm d}\theta}{2\pi} {\rm e}^{-{\rm i}N \theta - \beta 
\Psi(T, {\rm i}\theta/\beta)}
\nonumber \\
& = & \sqrt{\frac{\beta}{2L}}\exp \left(\frac{\pi L}{6 \beta} - \frac{1}{2} \pi L \beta \rho^2 \right).
\label{B7}
\end{eqnarray}
In the thermodynamic limit, the pre-factor is irrelevant when taking the log and we confirm Eq.~(\ref{B3}) for the
free energy density.
Since we are dealing with a Gaussian integral, we would get the same answer with the saddle point
method. However, note that the saddle point is purely imaginary,
\begin{equation}
\theta_0 = - {\rm i} \pi \beta N/L.
\label{B8}
\end{equation}
Since imaginary $\theta$ is tantamount to real chemical potential, this renders the method useless for lattice
calculations, at least for large volumes. 

We now illustrate the second physical interpretation of imaginary chemical potential mentioned above,
based on swapping (Euclidean) time and space. We compute the Casimir energy in an interval of length $L$
with quasi-periodic boundary conditions [cf. Eq.~(\ref{A7})] for massless fermions.
The Dirac spectrum is determined by the momenta discretized as
\begin{equation}
k_n = \frac{2\pi}{L}\left(n+\frac{1}{2}+\frac{\theta}{2\pi}\right).
\label{B9}
\end{equation}
The vacuum energy density, using heat kernel regularization,
\begin{equation}
{\cal E} = - \frac{1}{L} \sum_n |k_n| {\rm e}^{- \lambda |k_n|}
\label{B10}
\end{equation}
can now easily be worked out. Assuming  $\theta \in [-\pi,\pi]$ and taking the limit $\lambda \to 0^+$, one finds
\begin{equation}
{\cal E} = - \frac{1}{\pi \lambda^2} - \frac{\pi}{6 L^2} + \frac{\theta^2}{2\pi L^2}.
\label{B11}
\end{equation} 
Upon dropping the irrelevant quadratic divergence and  replacing $L$ by $\beta$, we indeed recover the
gc potential density $\psi$ at imaginary chemical potential, Eq. (\ref{B6}).

The case of massless fermions in 1+1 dimensions is instructive in the sense that the relationship between real
and imaginary chemical potential can be exhibited in closed form and related to known
analytic functions. Besides, it is important for our case study since it describes
the chirally symmetric phase of the Gross-Neveu model reached at high temperatures or chemical potential.
We shall come back to this point in Sect.~IV.

\section{Free massive fermions in 1+1 dimensions}
We next turn to free massive fermions in two dimensions. 
Dropping again $T$-independent divergent terms, the gc potential density is given by
an expression similar to Eq.~(\ref{B1}),
\begin{equation}
\psi=- \frac{1}{\beta \pi} \int_0^{\infty} {\rm d}k \ln \left[ \left(1+{\rm e}^{-\beta(\epsilon-\mu)} \right)
 \left( 1+{\rm e}^{-\beta(\epsilon+\mu)}\right) \right],
\label{C1}
\end{equation}
except that the dispersion relation is now $\epsilon=\sqrt{m^2+k^2}$.
At imaginary chemical potential, the argument of the log can be expanded and the momentum integration 
carried out with the help of the integral representation of a Bessel function,
\begin{eqnarray}
\psi &=& - \frac{1}{\beta \pi} \int_0^{\infty} {\rm d}k \ln \left[ \left(1+{\rm e}^{-\beta\epsilon+{\rm i}\theta} \right)
\left( 1+{\rm e}^{-\beta \epsilon-{\rm i}\theta}\right) \right]
\nonumber \\
& = & \frac{2m}{\beta \pi}\sum_{n=1}^{\infty} \frac{(-1)^n}{n}K_1(n\beta m)\cos (n\theta).
\label{C2}
\end{eqnarray}
The curves corresponding to Eqs.~(\ref{C1},\ref{C2}) look qualitatively like Fig.~1 and will not be shown here.
The only notable difference is the fact that the peaks in the upper curve are washed out since the singularities
move away from the imaginary $\mu$-axis. 
The Fourier series (\ref{C2}) converges at real but not at imaginary $\theta$, so that
it cannot be used directly for the analytic continuation (see however \cite{12a}). $\psi$ is an even, periodic function in $\theta$.
In the complex $\nu(=\beta \mu={\rm i}\theta)$ plane, the singularities are produced by the 
branch point of the log (vanishing of the argument), see Fig.~2 (due to the periodicity along the Im $\nu$
axis, a stripe of width 2$\pi$ contains all the information). 
\begin{figure}
\epsfig{file=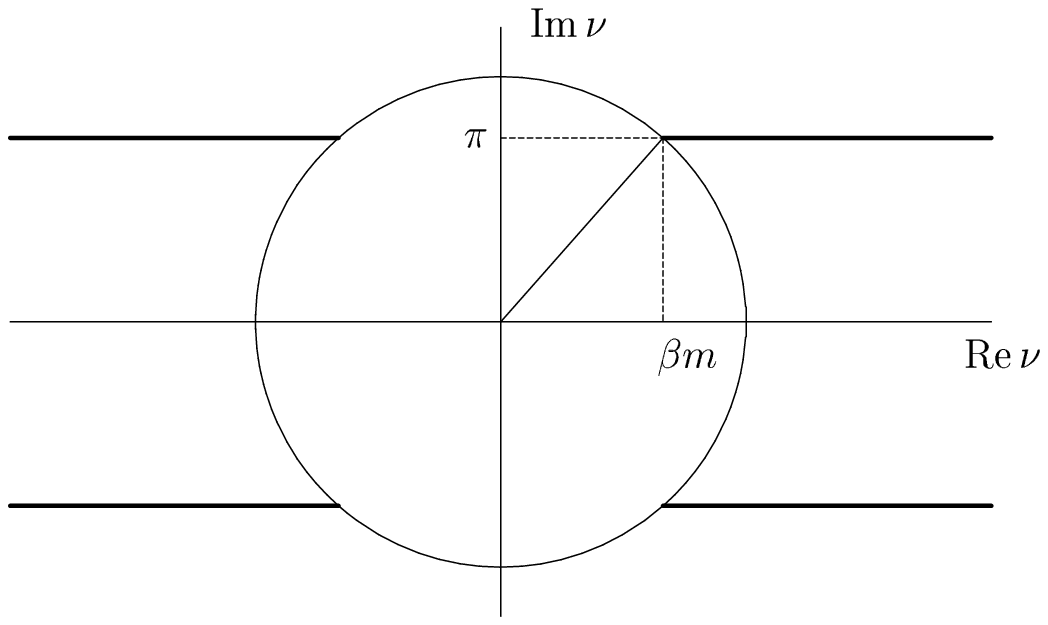,width=4cm,angle=270}
\caption{Cut structure (thick lines) of the grand canonical potential density $\psi$ for massive fermions
in the complex $\nu(=\beta \mu)$ plane. The circle delimits the
region of convergence of a series expansion around $\nu=0$.}
\end{figure}
A power series around $\nu=0$ has a radius of convergence given by the distance to the closest singularity,
\begin{equation}
\nu_{\rm max} = \sqrt{\beta^2 m^2 + \pi^2}.
\label{C3}
\end{equation}
For $m=0$ or high temperatures, we recover the value of $\pi$ from the previous section.
For large masses or low temperatures on the other hand, $\nu_{\rm max}$ is 
determined by $\beta m \gg \pi$ and therefore  much larger.
This will turn out to be important for the GN model later on. To illustrate the role of $\nu_{\rm max}$, we show 
a numerical example in Fig.~3. We compare the full gc potential at real $\nu$ with its power series expansion
around $\nu=0$, as it would be relevant for an analytic continuation from imaginary to real chemical
potential. The value of $\nu_{\rm max}$ clearly limits the convergence of the power series.  
\begin{figure}
\epsfig{file=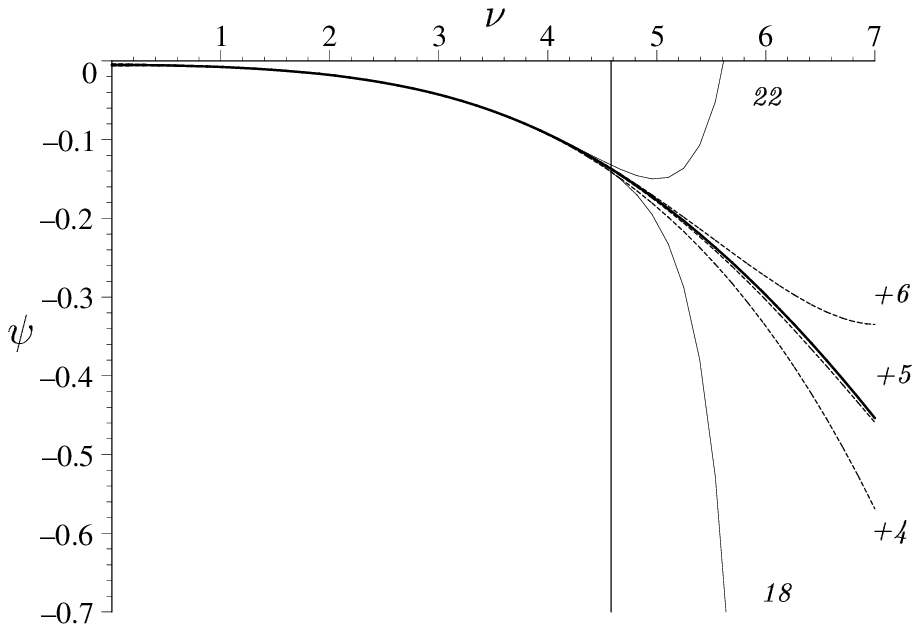,width=5cm,angle=270}
\caption{Grand canonical potential density for massive fermions at $T=0.3,m=1$, compared to various power
series expansions. Thin vertical straight line: $\nu_{\rm max}=4.58$, thick line: full calculation, thin lines: expansion
around $\nu=0$ to order $\nu^{18}$ and $\nu^{22}$, dashed lines: primary expansion around $\nu=0$
to order $\nu^{24}$, followed by secondary expansion around $\nu_0=2.29$ to 4th, 5th and 6th order in
$(\nu-\nu^0)$. The 5th order curve is almost indistinguishable from the full calculation on this plot.}
\end{figure}
In principle, one can continue analytically to any value of $\nu$ by a succession of expansions
around different points with overlapping circles of convergence.
By way of example, we first continue analytically to $\nu=\nu_0<\nu_{\rm max}$,
expanding around 0. The series expansion in $\nu_0$ has the radius of convergence again given
by the nearest singularity of $\psi$, i.e., allows us to go along the real axis up to
\begin{equation}
\nu_{\rm max} = \left\{ \begin{array}{ll}   \nu_0 +\sqrt{(\beta m-\nu_0)^2+\pi^2} & \mbox{if $0\le \nu_0\le \beta m$} \\
 \nu_0 + \pi  &  \mbox{if $\beta m \le \nu_0 \le \pi$} \end{array} \right.  
\label{C4}
\end{equation}
In practice, this requires a high order calculation in the primary expansion to get sufficient accuracy
to lower order in the secondary one (one has to find a kind of intermediate asymptotics). An example
is also included in Fig~3. Due to the high order in the expansion around $\nu=0$ needed, this would not
be a very realistic option in lattice QCD.

\section{Gross-Neveu model in 1+1 dimensions, assuming unbroken translational invariance}
The GN model is a four-fermion theory defined by the Lagrangian \cite{11}
\begin{equation}
{\cal L}= \sum_{i=1}^N \bar{\psi}^{(i)} {\rm i} \partial \!\!\!/ \psi^{(i)}+ \frac{1}{2}g^2\left( \sum_{i=1}^N
\bar{\psi}^{(i)}\psi^{(i)}\right)^2.
\label{D1}
\end{equation}
In the 't~Hooft limit ($N\to \infty$, $Ng^2=$ const.) its phase diagram is well understood.
Originally, translational invariance had been assumed at all temperatures and chemical potentials \cite{13}. Then
the only issue is the fate of the discrete chiral symmetry $\psi \to \gamma_5 \psi$. It is broken in the vacuum
(leading to a dynamical fermion mass) but gets restored at high $T$ or $\mu$, see Fig.~4. 
\begin{figure}
\epsfig{file=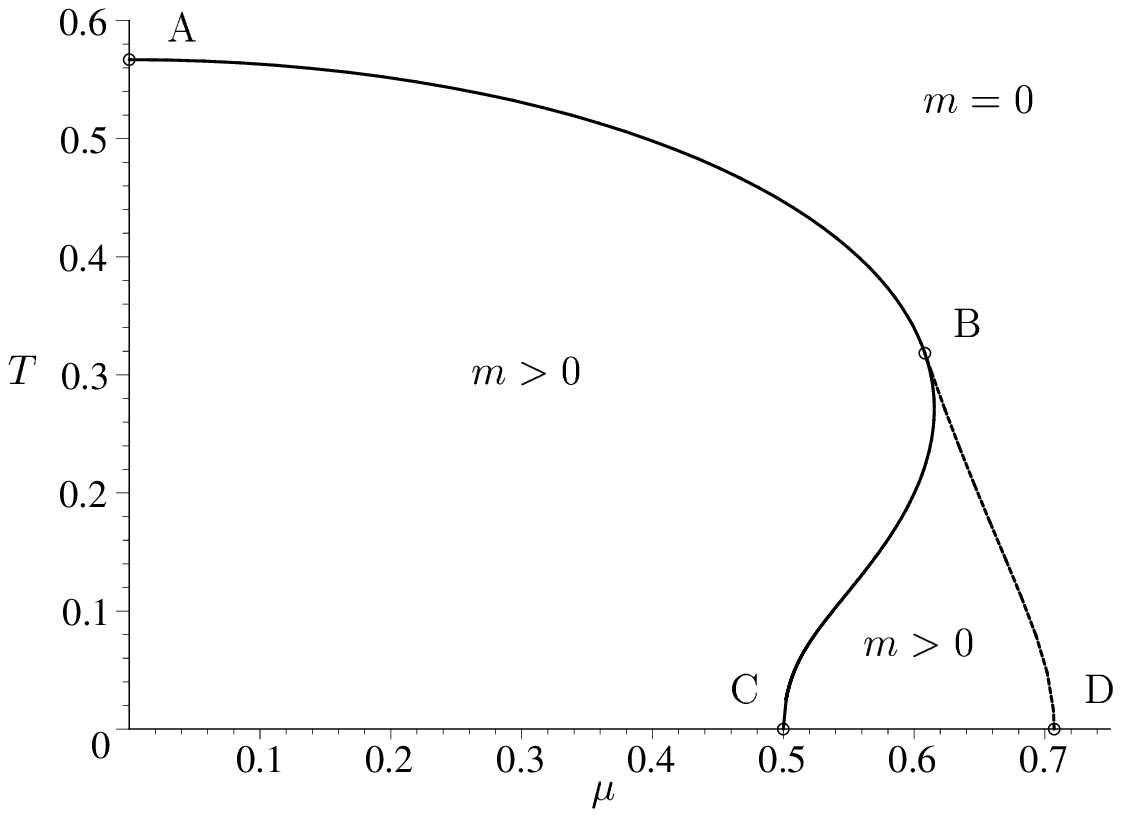,width=5cm,angle=270}
\caption{Phase diagram of Gross-Neveu model at real chemical potential, assuming unbroken translational
invariance \cite{13}.
AB is a 2nd order line, BD a first order line, B a tricritical point. In the region BCD the gc potential has
two local minima, the global minimum being the one at $m>0$.}
\end{figure}
The tricritical point B where the 2nd order line AB goes over into a first order line BD  is reminiscent of QCD.
More recently, a third, inhomogeneous phase (a kink-antikink crystal) has been identified \cite{14,15}. It is more stable at
high $\mu$, low $T$ and requires a revision of the phase diagram shown in Fig.~4, cf. Sect.~V.
In the present section, we take as our model the GN model with homogeneous phases only for the sake 
of simplicity. We shall return to the crystal phase in the next section. The question we ask ourselves is:
Suppose we were able to solve the GN model at imaginary $\mu$ only, although to any desired precision.
To what extent could we reconstruct the phase diagram in Fig.~4 from such data?

The solution of the GN model is well documented in the literature and does not have to be repeated here (see \cite{16}
and references therein).
Since at large $N$ the Hartree-Fock approximation becomes exact and the scalar mean field acts like a mass, 
the gc potential differs from the one for free, massive fermions of Sect.~III only through a double counting
correction to the potential energy. An important novel aspect is self-consistency: The mass is no longer
an external parameter but determined dynamically, namely by minimization of $\psi$ through the gap equation. 
In units where the vacuum fermion mass is set equal to 1, the renormalized gc potential density reads 
\begin{equation}
\psi = \frac{m^2}{4\pi}\ln m^2 + \frac{1-m^2}{4\pi} + \psi_{\rm free}(m)
\label{D2}
\end{equation}
Here, $\psi_{\rm free}$ is given in Eqs. (\ref{C1},\ref{C2}) at real and imaginary chemical
potential, respectively. Minimization of $\psi$ with respect to $m$ at real $\mu$ results in the
phase diagram of Fig.~4.

Let us first mimic the procedure which has been applied in QCD lattice calculations at imaginary potential.
We minimize the gc potential (\ref{D2}) at imaginary chemical potential with respect to $m$ and find the
phase diagram in Fig.~5. Since $\psi$ is periodic in $\theta$, it is sufficient to display one period, say $-\pi<\theta\le \pi$.
\begin{figure}
\epsfig{file=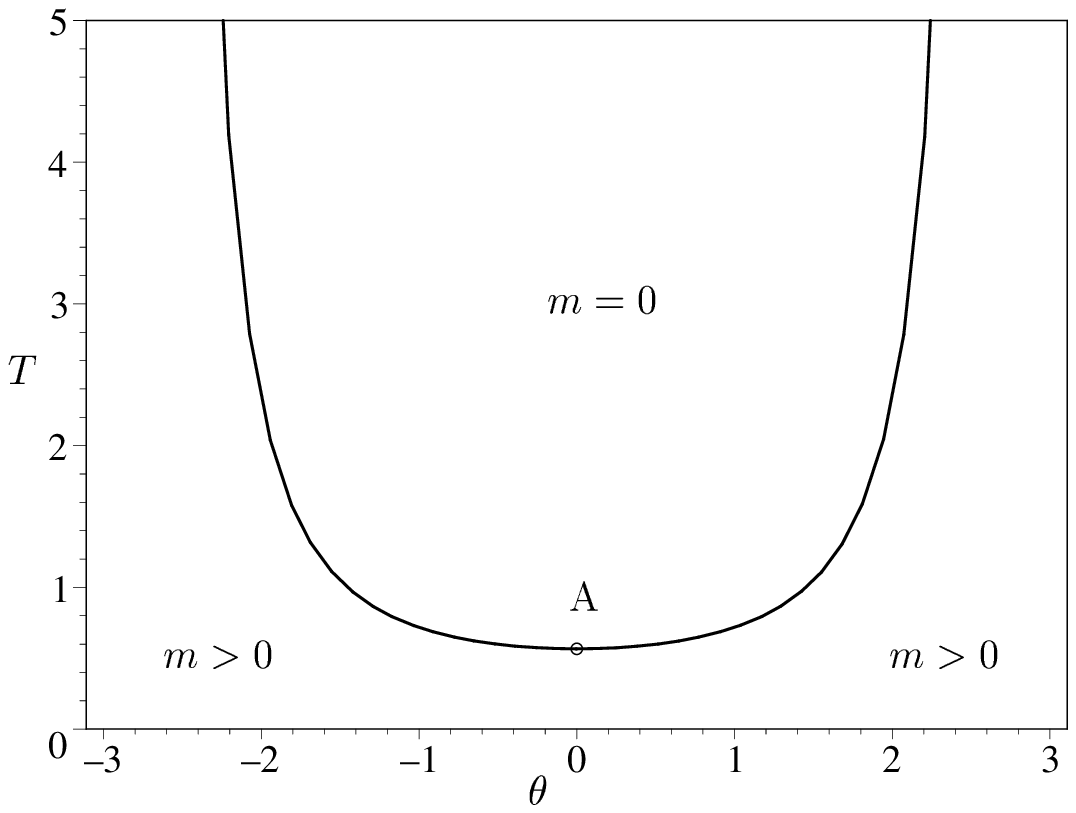,width=5cm,angle=270}
\caption{Phase diagram of Gross-Neveu model at imaginary chemical potential $\mu={\rm i}\theta/\beta$,
assuming unbroken translational invariance. This figure should be periodically continued to the left and
to the right. The phase boundary is a 2nd order line, the analytic continuation of curve ABC in Fig.~4 (the point A
is common). The lines $\theta=\pm \pi$ are asymptotes to the phase boundary.}
\end{figure}
Inspection of the effective potential shows that the curve is a 2nd order phase
boundary where the dynamical fermion mass vanishes continuously. 
Figs.~4 and 5 do not have much resemblance. They only coincide at $\mu=\theta=0$, see the common point A
on the phase boundaries. 

Before turning to the relationship between the two phase diagrams
via analytic continuation, let us try to understand the physics behind the unfamiliar result 
at imaginary chemical, Fig~5. For this purpose, it is useful to return to the interpretation of imaginary
$\mu$ in terms of the temperature inversion symmetry discussed in the Introduction. 
If we adopt the picture of the Casimir effect, the phase diagram in Fig.~5 tells us that for a given
boundary condition phase $\theta$ (or, equivalently, a magnetic field in an extra dimension), there is always
a 2nd order quantum phase transition with restoration of chiral symmetry if we decrease the size of the
compact space direction. The critical length $L$ is the inverse of the critical temperature $T$ plotted in Fig.~5: 
It decreases with increasing $\theta$ and vanishes at the bosonic point $\theta=\pi$, i.e., for periodic boundary conditions
(this corresponds to the asymptotes in Fig.~5). This claim can easily be made more precise.
In Sect. 4.2 of \cite{16}, the gap equation for the GN model on a finite interval with anti-periodic
boundary conditions was derived. Setting $m=0$, the critical length or,
by temperature inversion symmetry, the critical temperature at $\mu=0$ was found. It is a simple
exercise to repeat this derivation with quasi-periodic boundary conditions, resulting in the condition
\begin{equation}
\ln \Lambda = \sum_{n=0}^{N_{\Lambda}} \left( \frac{1}{2n+1+\theta/\pi}+ \frac{1}{2n+1-\theta/\pi}\right)
\label{D3}
\end{equation}
with $N_{\Lambda}=L\Lambda/(4\pi)$. The sums can be performed in terms of the 
digamma function $\Psi(z)=(\ln \Gamma(z))'$ and yield in the limit $\Lambda\to \infty$
\begin{equation}
2 \ln \left(\frac{4\pi}{L}\right) +  \Psi\left( \frac{1}{2}- \frac{\theta}{2\pi} \right)
+  \Psi\left( \frac{1}{2}+ \frac{\theta}{2\pi} \right)=0.
\label{D4}
\end{equation}
Replacing $L$ by $\beta=1/T$, this is indeed the implicit equation describing the true phase boundary in Fig.~5,
now interpreted as a quantum phase transition when compressing fermions at $T=0$ for different boundary conditions
parametrized by $\theta$.

We now go back to the phase diagram at real $\mu$ in Fig.~4 and the relation between these two
phase diagrams. At real chemical potential, the 2nd order line ABC in Fig.~4 has been computed
analytically using perturbation theory some time ago with the result \cite{17,15}
\begin{equation}
2 \ln \left(\frac{4\pi}{\beta}\right) +  \Psi\left( \frac{1}{2}+ \frac{{\rm i}\nu}{2\pi} \right)
+  \Psi\left( \frac{1}{2}- \frac{{\rm i}\nu}{2\pi} \right)=0.
\label{D5}
\end{equation}
This agrees perfectly with Eq.~(\ref{D4}) once we replace $\nu$ by ${\rm i}\theta$ and identify $\beta$ with $L$.
These curves can therefore be related by analytic continuation. 

Following the method which has been applied to QCD, we try to infer the phase boundary at real $\mu$
from the one at imaginary $\mu$ by a power series around $\mu=0$.
The digamma function $\Psi(z)$ is an analytic function and has the nearest singularity at $z=0$.
The radius of convergence of the power series of Eq.~(\ref{D4}) is therefore $\nu=\pi$.
Since the fermion mass vanishes at the boundary, it is not surprising that we recover the 
same value which we encountered in the expansion of  the gc potential for free massless fermions in Sec.~II.
In Fig.~6, we show several approximations 
corresponding to different orders in $\nu$. With a sufficiently large number of terms one gets
an accurate picture of the true phase boundary AB, but one can also continue past the tricritical
point B where the curve ceases to be the phase boundary. By the method of overlapping circles
of convergence, one would be able even to go beyond $\nu=\pi$, as is clear from the analytic
structure of Eq.~(\ref{D4}).
\begin{figure}
\epsfig{file=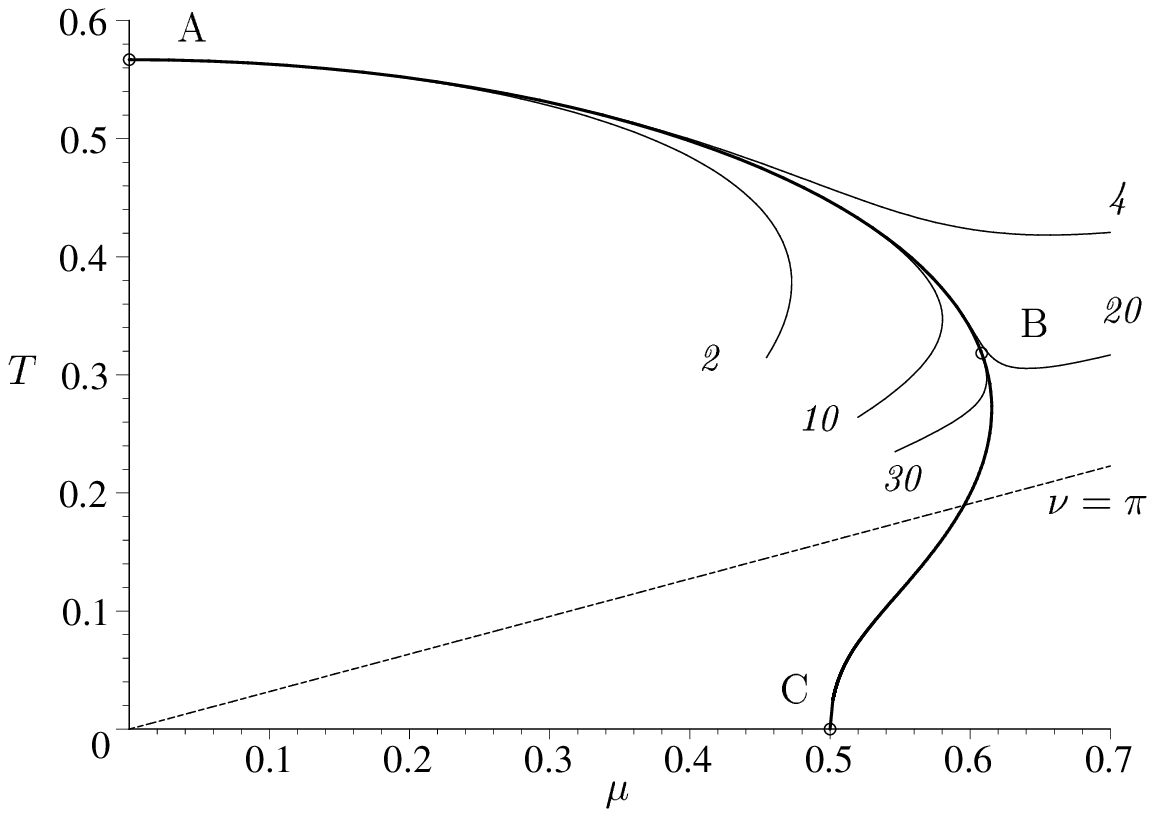,width=5cm,angle=270}
\caption{2nd order critical line in comparison to analytic continuation from imaginary to real $\mu$ 
via power series expansion around $\theta=0$. The curves correspond to 2nd, 4th, 10th, 20th, 30th order
in $\theta$ and reach beyond the tricritical point B. The series converges above the line $\nu=\pi$.}
\end{figure}
Going back to the full phase diagram in Fig.~4, we observe that the calculation at imaginary $\mu$
followed by an arbitraryly precise analytic continuation reproduces the 2nd order line but gives
no clue whatsoever about the location of the tricritical point B, let alone the first order line BD.

Where has the information about the first order phase transition gone in the process of making $\mu$ 
imaginary? The reason for this apparent loss of information is clear: The phase structure is contained
in the shape of the effective potential $\Psi$ as a function of $m$, for given $T,\mu$. Thus for example,
along the first order line, $\Psi$ has two local minima of equal depth. In the present case, 
the potential at imaginary $\mu$ has less structure and displays only one minimum. This 
may be caused by a change of sign of some coefficient in the series expansion in $m$
under the transition $\nu \to {\rm i}\theta$. If we only consider the minima of the respective potentials,
there is simply no way to recover the full information. This is exactly what happens if one performs the
thermodynamic calculation at imaginary $\mu$ as above. 
What one evidently has to do is go back one step from the observables (computed in the minimum)
to the full $m$-dependent effective potential. In other words, we should keep the order parameter fixed
and do the analytic continuation not just once, but for a whole range of $m$-values. In this way one
might hope to reconstruct the effective potential at real $\mu$ from that computed at imaginary $\mu$
and then infer the phase diagram as usual.

Since the $\mu$-dependent part of the gc potential is the same as for free, massive fermions, we have
all the necessary ingredients already. The crucial question about the convergence of the power series
of $\Psi$ in $\nu$ can be answered with the help of Sect.~III. By way of example, consider the point $T=0.15,\mu=0.66$
close to first order line. Eq.~(\ref{C4}) yields a lower bound for the mass variable for which the series
around $\theta=0$ converges, namely 
\begin{equation}
m>\sqrt{\mu^2-\pi^2 T^2}
\label{D6}
\end{equation}
or $m>0.46\ (\beta m >3.08)$ in the case at hand. This can be verified, see Fig.~7 for an illustration. 
\begin{figure}
\epsfig{file=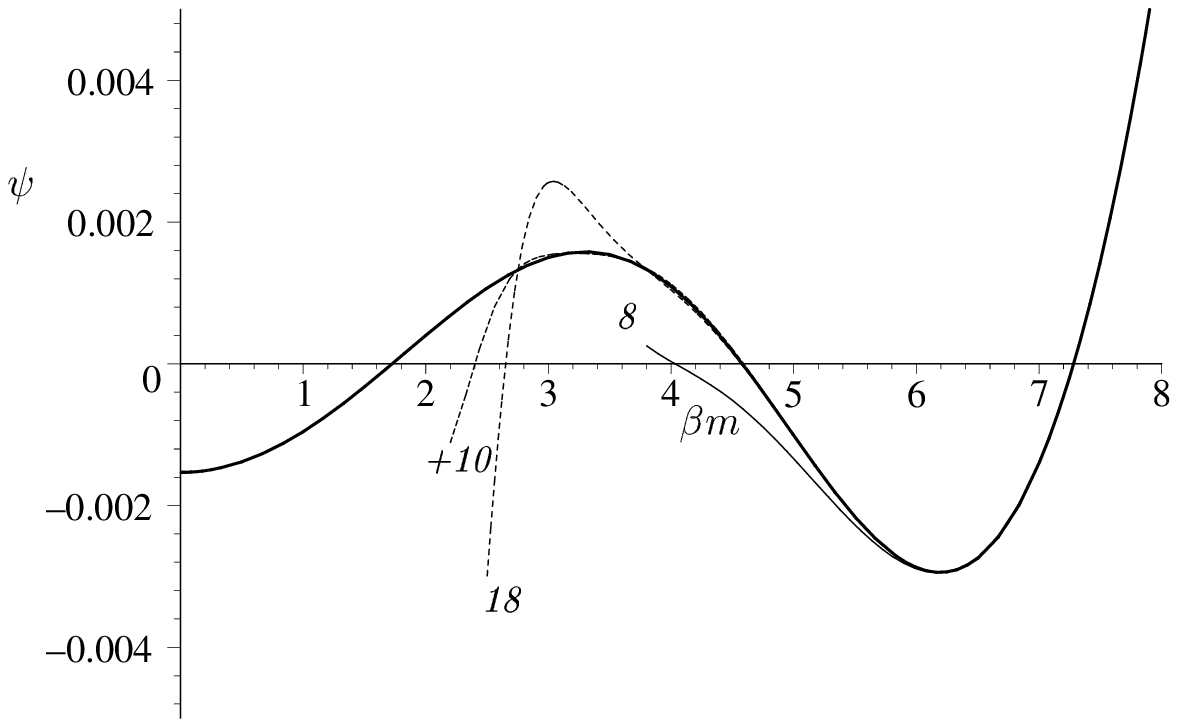,width=5cm,angle=270}
\caption{Mass dependent effective potential of the GN model at $T=0.15,\mu=0.66$. Thick line: full
calculation, thin solid line: analytic continuation via series expansion around $\theta=0$, order $\theta^8$,
dotted line: dto., order $\theta^{18}$, dash-dotted line: order $\theta^{18}$ expansion around $\theta=0$ 
followed by order $(\theta-\theta_0)^{10}$ expansion around $\theta_0=\pi/2$.}
\end{figure}
By successive analytic continuations around different points, the region can be expanded as pointed out
above, at the cost of high order calculations. An example is included in Fig.~7.

In this particular example, although we would not be able to determine the full $m$-dependence
of the effective potential, the non-trivial minimum can be reconstructed with a reasonable effort.
It is well within the region of convergence of the series around $\theta=0$. On the other hand, the
value in the trivial minimum ($m=0$) could be obtained independently following Sect.~II, so that 
one could decide which minimum is deeper. This is more generally true at low temperatures:
The bound in Eq.~(\ref{C3}) is dominated by the mass term, since the mass at the minimum is close to 1.
The most difficult region is actually a small window in the vicinity of $\nu=\pi$, where the non-trivial minimum
violates the bound (\ref{C3}), at least in the truncation of the power series we are using. 
Thus, provided one has a rough idea of the shape of the effective potential, one can actually 
reconstruct the GN phase diagram, Fig.~4, in full glory, starting from a calculation of the $m$-dependent 
effective potential at imaginary chemical potential. This includes the information about the order of 
the phase transition. In Fig.~8,  we show the result of such a calculation, leaving out the above mentioned
window where our method fails at the present level of accuracy. 
The tricritical point has been located and the unphysical branch BC of the 2nd order critical line (see Fig.~6)
has disappeared.
\begin{figure}
\epsfig{file=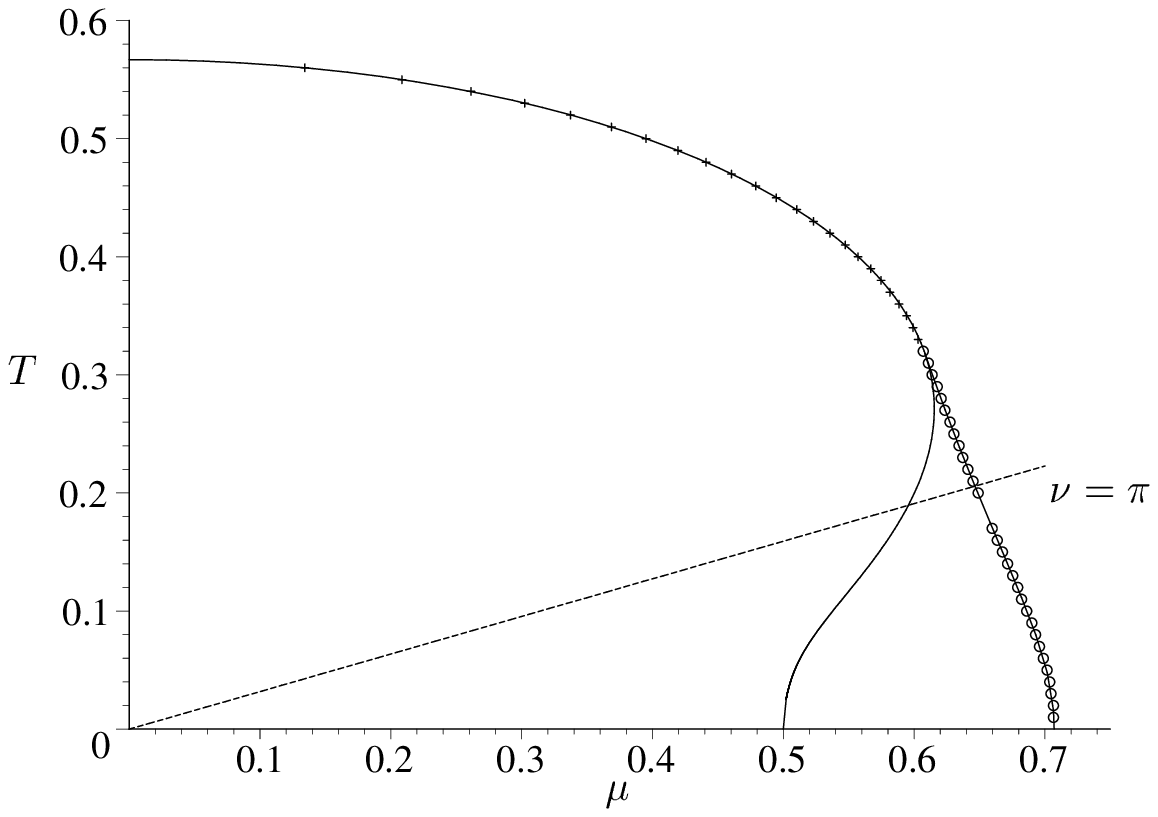,width=5cm,angle=270}
\caption{Final results for reconstructed phase boundaries in the GN model, using only imaginary $\mu$
input and power series expansion to order $\theta^{10}$.  The crosses are identified as 2nd order, the
circles as first order phase transition. The gap in the circles below $\nu=\pi$ is the region where the
accuracy was not enough to get reliable results. The full lines are the exact results.}
\end{figure}

\section{Kink-antikink crystal from imaginary chemical potential?}

The phase diagrams in Fig.~4 and Fig.~5 have been derived under the assumption that the order parameter
$\langle \bar{\psi}\psi \rangle$ is constant in space, i.e., the mean field acts like a mass term. Actually,  at real
chemical potential, an inhomogeneous phase with a periodically modulated
scalar condensate, a kink-antikink crystal, is more stable in some part of the $(\mu,T)$-plane  \cite{14,15}.
The phase diagram of the GN model including the crystal phase is shown in Fig.~9. 
\begin{figure}
\epsfig{file=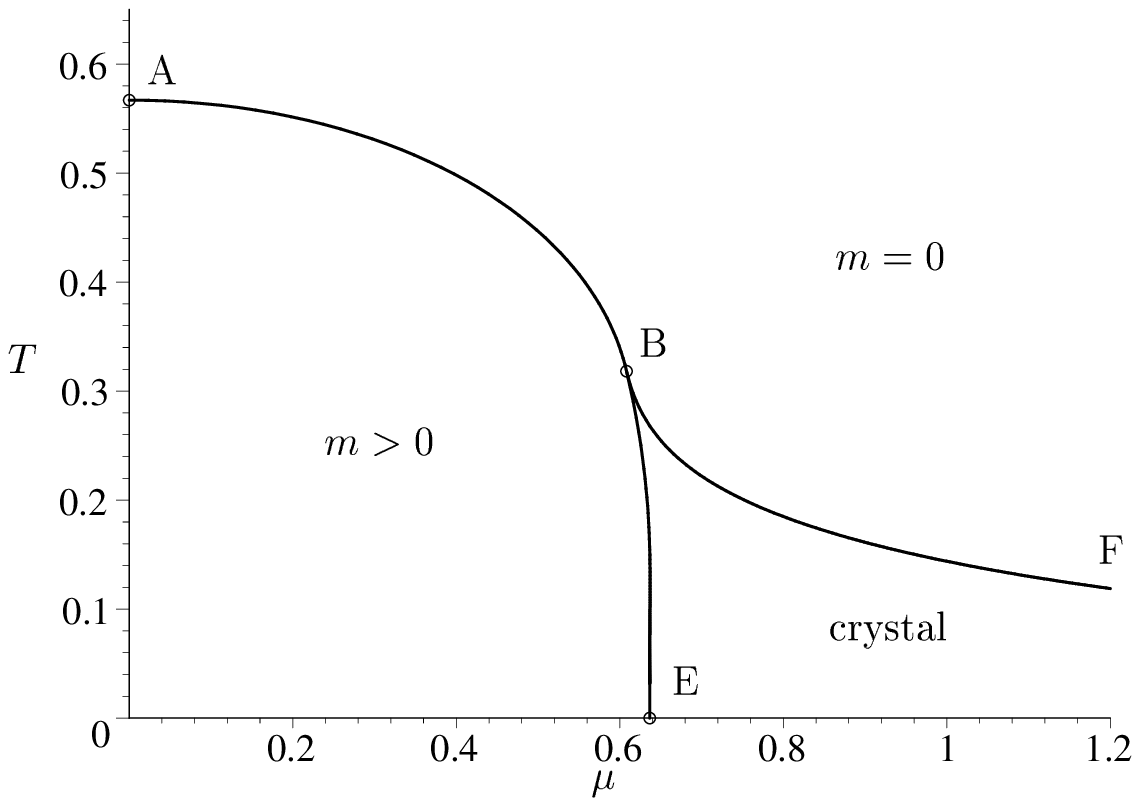,width=5cm,angle=270}
\caption{Full phase diagram of the GN model at real chemical potential \cite{14,15}. The critical line AB is the same 
as in Fig.~4, the point E is at $T=0,\mu=2/\pi$ (the baryon mass), the novel kink-antikink crystal is bounded by the 
second order lines BE and BF. In this region, the scalar condensate is inhomogeneous and given by an
elliptic function, see Eq.~(\ref{E1}).}
\end{figure}
This immediately raises the question: How would the phase diagram
at imaginary chemical potential change if we would allow for inhomogeneous mean fields? Is it at
all possible to infer the existence and properties of the crystal phase from a computation at imaginary
$\mu$? 

We first argue that it is very unlikely that an inhomogeneous phase would be favoured at imaginary
$\mu$ and any temperature. Physically, at real $\mu$ it arises as a consequence of the Peierls instability
in one-dimensional fermion systems, i.e., dynamical creation of a gap at the Fermi surface \cite{18}. To see
what happens at 
imaginary $\mu$, it is again advantageous to switch to the Casimir interpretation: $T,\mu$ are mapped onto  
the size $L$ and the boundary phase $\theta$, and there is no known mechanism which would
induce a breakdown of translational invariance as a function of these parameters in a quantum phase 
transition. To further test this intuitive expectation, we have repeated the calculation of the gc potential
for the GN model, using a two-parameter family of periodic scalar potentials which contain the self-consistent
one at real chemical potential, namely the elliptic function
\begin{equation}
S(x)= A \kappa^2 \frac{{\rm sn}(Ax,\kappa) {\rm cn}(Ax,\kappa)}{{\rm dn}(Ax,\kappa)}.
\label{E1}
\end{equation}
Since the calculation is analytical up to one-dimensional numerical integrations, we were able to switch from real
to imaginary chemical potential but found no non-trivial minimum in the imaginary case. 
This does not rule out periodic potentials of different shape. The following argument shows that there is at least no
perturbative instability towards breaking translational invariance: Using almost degenerate perturbation
theory, one can derive the phase boundary ABF in Fig.~9 without reference to the exact potential, Eq.~(\ref{E1}).
The result at real chemical potential is encoded in the equation \cite{15}
\begin{eqnarray}
4 \ln \left(\frac{\beta}{4\pi}\right) & = & \min_{a\ge 0}\left[ \Psi \left( \frac{1}{2} + \frac{{\rm i}(\nu+a)}{2\pi}\right)
\right.
\label{E2} \\
& & + \left. \Psi \left( \frac{1}{2} - \frac{{\rm i}(\nu+a)}{2\pi}\right)
+ (a \to -a) \right].
\nonumber
\end{eqnarray} 
At real $\nu$, the minimum is either at $a=0$ (corresponding to the boundary AB of the homogeneous phase)
or at $a>0$ (the boundary BF of the crystal phase). If we switch to imaginary $\nu$, the non-trivial minimum
disappears. For $a=0, \nu={\rm i}\theta, \beta=L$ Eq.~(\ref{E2}) leads us back to Eq.~(\ref{D4}) above so that
there is no indication of any inhomogeneous phase at imaginary $\mu$.

Following the reasoning of the last section, it is then clear that one should not think of analytically continuing
observables or phase boundaries, but should go back one step. Whereas we dealt with the $m$-dependent
effective potential there, now we have to consider the full effective action, a functional of the $x$-dependent 
scalar mean field $S(x)$. The idea would be to compute the effective action (i.e., the gc potential density) at
imaginary $\mu$ and do the analytic continuation to real $\mu$ before minimization. In practice, one could restrict
oneself to a few
parameter family of functions $S(x)$, so that the effective action would become a function of these parameters,
thereby generalizing the function of one parameter $m$ in the translationally invariant case.
We have not done this calculation for the GN model. However we can illustrate the idea using Eq.~(\ref{E2})
for the phase boundary where the analytic structure is fully laid out. Let us pretend that we could derive Eq.~(\ref{E2})
at imaginary chemical potential only. Then we would not find any non-trivial minimum, as mentioned above. 
However we could analytically continue the r.h.s. via a power series expansion in $\nu$ and check the 
accuracy needed in order to see the non-trivial minimum, i.e., the instability with respect to crystallization.
The result of such a calculation is shown in Fig.~10. 
\begin{figure}
\epsfig{file=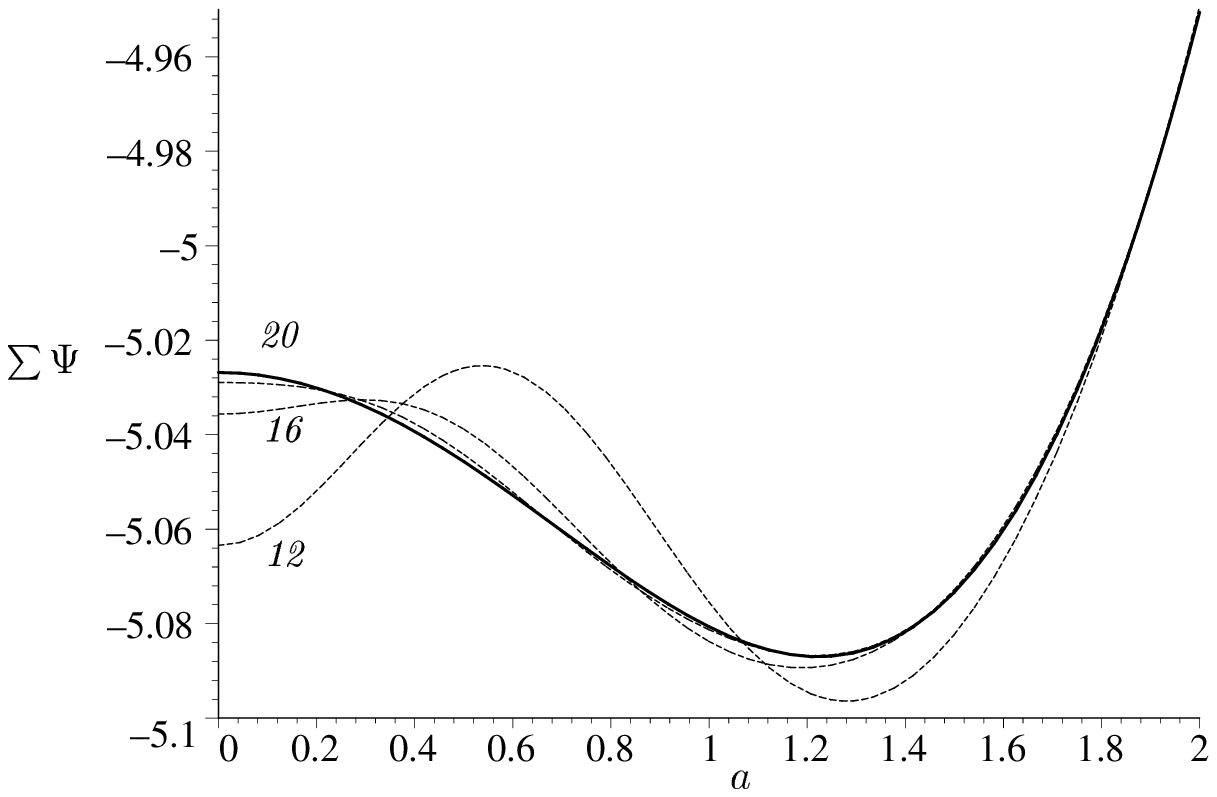,width=5cm,angle=270}
\caption{Expression in square brackets on the right hand side of Eq.~(\ref{E2}) for $\nu=2.2$.
Solid curve: full calculation, dashed curves: analytic
continuation via power series expansion around $\theta=0$ to order $\theta^{12},\theta^{16},\theta^{20}$.}
\end{figure}
Although the required order in the series expansion is quite high, it is interesting that there is a method which in
principle allows us to infer the existence of a crystal phase, using only imaginary $\mu$ input. Incidentally, the radius of
convergence as given by nearest by singularity of the digamma functions is $\nu=\sqrt{\pi^2+a^2}$,
whereas the minimum for large $\nu$ is located close to $a=\nu$. Hence there is no limitation to $\nu=\pi$ in
this particular calculation and no fundamental obstacle to derive the phase boundary for large $\mu$ and
low temperatures, far beyond the tricritical point.

\section{Summary and conclusions}
In this work, we have studied the relationship between a relativistic quantum field theory at real and imaginary
chemical potential. Since lattice QCD can only be simulated at imaginary $\mu$ using the standard
Monte Carlo method, we were interested in how to retrieve thermodynamic information from imaginary $\mu$.
Using an exactly solvable toy model, the (large $N$) GN model in 1+1 dimensions, we first
recovered the well-known limitations met if one tries to analytically continue phase boundaries or other
thermodynamic observables. In order to overcome these barriers, one has to go one step back in the
formalism and compute the full effective potential of the order parameter (here, the chiral condensate $\bar{\psi}\psi$)
at imaginary $\mu$, before minimization. In the GN model studied here, this enabled us to reconstruct the
full phase diagram with homogeneous phases, including the location of the 
tricritical point and the first order phase boundary. A similar computation of an effective action instead of an effective
potential is likely to yield information about inhomogeneous phases. 

It is possible that these results carry over to lattice QCD as well. 
One would have to compute the effective potential (for homogeneous phases) or the effective 
action (for inhomogeneous phases) at imaginary chemical potential. Although we do not have the
expertise to judge whether such calculations are feasible with Monte Carlo methods, it may be
worthwhile to think along these lines to learn more about the QCD phase diagram from first principles.

We should like to thank Frieder Lenz and Oliver Schnetz for useful remarks and Maria Paola Lombardo, 
Massimo D'Elia and Philippe de Forcrand for helpful comments on the first version of this paper.

\end{document}